\documentclass[reprint, longbibliography, amsmath,amssymb, aps, superscriptaddress]{revtex4-1}

\usepackage{graphicx}
\usepackage{textcomp}

\usepackage{dcolumn}
\usepackage{bm}
\usepackage{color}
\usepackage{soul}
\usepackage{hyperref}
\usepackage{float}
\usepackage{amsmath}
\usepackage{multirow}
\usepackage{ulem}
\begin{document}
	
	\newcommand{\Pc}{H$\mathrm{_2}$Pc~}
	\newcommand{\dIdV}{d$I$/d$V$~}
	
	\newcommand{\change}[2]{#2}
	
	\title{Tuning spinaron and Kondo resonances via quantum confinement}
	\author{Markus Aapro}
	\email[Corresponding authors. ]{markus.aapro@aalto.fi, abraham.kipnis@aalto.fi, peter.liljeroth@aalto.fi}
	\affiliation{Department of Applied Physics, Aalto University, Finland}

	\author{Abraham Kipnis}
	\email[Corresponding authors. ]{markus.aapro@aalto.fi, abraham.kipnis@aalto.fi, peter.liljeroth@aalto.fi}
	\affiliation{Department of Neuroscience and Biomedical Engineering, Aalto University, Finland}

	\author{Jose L. Lado}
	\affiliation{Department of Applied Physics, Aalto University, Finland}
	
	\author{Shawulienu Kezilebieke}
	\affiliation{Department of Physics, Department of Chemistry and Nanoscience Center, 
		University of Jyväskyl\"a, FI-40014 University of Jyväskyl\"a, Finland}
	
	\author{Peter Liljeroth}
	\email[Corresponding authors. ]{markus.aapro@aalto.fi, abraham.kipnis@aalto.fi, peter.liljeroth@aalto.fi}
	\affiliation{Department of Applied Physics, Aalto University, Finland}

	\begin{abstract}
		Controlling zero bias anomalies in magnetic atoms provides a promising strategy to engineer tunable quantum many-body excitations.
		Here we show how two different quantum impurities featuring spinaron and Kondo excitations can be controlled via quantum confinement engineering by using circular quantum corrals on a Ag(111) surface. In corrals built from both Ag and Co adatoms, the width of the zero bias anomaly in the central Co adatom oscillates as a function of corral radius with a period of half of the Ag(111) surface state wavelength. Parameters extracted for Co/Ag(111) show only small differences in extracted spinaron zero-bias anomaly between corral walls built from Ag or Co adatoms. In quantum corrals occupied with metal-free phthalocyanine, a \change{paradigmatic }{}$S=1/2$ Kondo system, we observe notable changes in the zero bias anomaly lineshape as a function of corral radius. Our results offer insight into many-body Kondo and spinaron resonances where the electronic density is controlled by confinement engineering.
	\end{abstract}
	\maketitle
	\section{Introduction}
	
	The Kondo effect occurs when magnetic impurities are introduced into a non-magnetic conductor \cite{Jun1964}. Local magnetic moments in the impurities interact with the conduction electrons in the non-magnetic host, resulting in a many-body spin singlet ground state and increasing resistance at low temperatures. The Kondo effect has been studied extensively in bulk and 2D materials and quantum dot devices \cite{Goldhaber-Gordon1998_SETKondo, Chang2009_QD_Kondo_review}. There is further interest in observing the Kondo effect in tunneling spectroscopy of single atomic and molecular impurities \cite{Madhavan1998-first-Kondo-STS,Nagaoka2002-Ti-Ag100,Ternes2009-single-atom-Kondo-review,Zhang2013a}.

	Low temperature scanning tunneling microscopy (STM) experiments have demonstrated precise manipulation and placement of surface adsorbates for investigating quantum effects in low dimensional systems.
	One of the first examples of this were quantum corrals \cite{Crommie1993}, nanostructures that confine surface state electrons to a two-dimensional geometry resulting in formation of confined modes. Their shapes and eigenenergies are determined by the corral shape and size, as well as the surface state dispersion. Engineering the confined modes offers various functionalities \cite{Li2020a-corral-review-APL} such as quantum phase extraction \cite{Moon2008a}, single-atom gating of eigenmode superpositions \cite{Ngo2017, Moon2008-single-atom-gating}, molecular adsorbate tautomerization and manipulation \cite{Kugel2017-atom-dropping,Hla2004-molecular-shooting}, and mirage effects \cite{Stepanyuk2005,Manoharan2000-mirage,Li2020-Kondo-free-mirage}. Surface state electrons confined to artificial lattices can inherit properties associated with the lattice geometry, such as non-integer dimensionality \cite{Kempkes2019_fractals}, topological states \cite{Kempkes2019_HOTI} and Dirac dispersion \cite{Gomes2012a, Ako2019_review}. 
	The interaction strength between magnetic impurities and electrons in noble metal surface states has been investigated \cite{Li2018a-corrals,Henzl2007,Limot2004-step-edge,Moro-Lagares2018-eF_mapping_Kondo, Schneider2005_CoKondo}, and quantum corrals emerge as a model platform for studying the Kondo effect and its interaction with surface states. Still, details on the coupling between surface states, confined modes and Kondo impurities remain unresolved. Quantum corrals on Ag(111) offer two benefits over those built on other noble metal surfaces: the confined mode energy widths are small and due to the surface state onset energy being close to the Fermi energy ($E_\mathrm{F}$), it is possible to build corrals with individual (or zero) occupied confined modes. 
	
	The cobalt adatom was the first Kondo-like impurity studied with scanning tunneling spectroscopy (STS) \cite{Madhavan1998-first-Kondo-STS}, and has since been studied on various substrates \cite{Knorr2002, Ren2014_Co_on_Graphene_Kondo}, in quantum corrals \cite{Manoharan2000-mirage, Li2018a-corrals}, atomic chains \cite{Moro-Lagares2019-Co_Ag111-chains, Foelsch2004_CoCuChains} and lattices \cite{Figgins2019-Kondo-lattices}. Many studies have discussed the role of the Ag(111) surface states in the zero bias anomaly: variations in low-energy spectra have been observed as a function of lateral distance from step edges \cite{Li2018a-corrals,Limot2004-step-edge}, distance from other adatoms \cite{Li2018a-corrals, Moro-Lagares2019-Co_Ag111-chains, Figgins2019-Kondo-lattices}, and confinement within quantum corrals and lattices \cite{Li2018a-corrals, Figgins2019-Kondo-lattices}. 
	Atomic manipulation was used \cite{Moro-Lagares2018-eF_mapping_Kondo} to map the surface electron density at 3 mV across Ag(111) free from impurities, move a Co adatom around the area and measure STS at each point to correlate the zero bias anomaly width to surface state density. Their work showed a positive correlation between these variables, suggesting that the zero-bias anomaly can be tuned via the surface state density, further suggesting that the system behaves like
	a two-channel SU(4) Anderson model rather than a one-channel SU(2) model \cite{Moro-Lagares2018-eF_mapping_Kondo}.
	It is also important to study how Kondo impurities with different spin states and Anderson model parameters \cite{Anderson1961_model} behave in quantum corrals: most studies on Ag(111) have focused only on the Co adatom, which has 
	multiple orbitals contributing to many-body scattering \cite{Valli2020_CoSpinOnCu}.

	Interestingly, despite its long-standing consideration as a Kondo state, recent studies on the Co/Ag(111) system suggest an alternative explanation for the zero bias anomaly \cite{Bouaziz2020-spinaron, friedrich2023spinaron,Noei2023}. These studies argue that it arises from gapped spin-excitations leading to a spinaron excitation
	. In contrast with conventional $S=1/2$ magnetic impurities, the dominant nature of spin excitation in Co stems from the strong magnetic anisotropy of the
	system due its higher spin.
	The spinaron is a many-body excitation associated with a magnetic polaron, which arises from the interplay between the spin excitation of the Co and
	the polarization of the metallic electron cloud.
	The renewed understanding of the Co zero bias anomaly motivates the development of strategies to tune and disentangle its microscopic nature, and specifically contrast it with well defined Kondo systems.

	To compare with the spin excitation and spinaron of Co, a pure $S=1/2$ magnetic impurity system would be highly desirable.
	Phthalocyanine (Pc) and its metal complexes have been widely used as individual magnetic impurities due to their chemically stable 2D structure. On adsorption of metal-free \Pc molecules on the Ag(111) surface, charge transfer causes the lowest unoccupied molecular orbital (LUMO) to become occupied. 
	Granet et al.~studied \Pc molecules in the dilute limit and in self-assembled monolayers on Ag(111) using STS and UPS, showing that \Pc 
	exhibits the Kondo effect on Ag(111) in both lattices and dilute films \cite{Granet2020}.
	The Kondo peak width, associated with the Kondo temperature of the system, was found to be smaller for \Pc molecules in self-assemblies as compared to isolated molecules. 
	Similar differences in magnetic behaviour have also been observed with \Pc on Pb(100), where a single pair of Yu-Shiba-Rusinov states emerges in supramolecular arrays of \Pc molecules when the LUMO becomes occupied \cite{Homberg_H2Pc_YSR_Pb}. The absence of \textit{d}- and \textit{f}-electrons makes the \Pc molecule a model system of $\pi$-magnetism, which has attracted intense research interest \cite{deOteyza2022_carbon_magnetism_review}. So far, no studies have reported using quantum corrals to explore how modifying the surface state density influences the Kondo effect in H$\mathrm{_2}$Pc/Ag(111).

	Here, we use quantum corrals to tune the coupling of Co adatoms and \Pc molecules with the Ag(111) surface state 
	and study the corresponding changes in the tunneling spectra to better understand the interactions between magnetic impurities and their electronic environment.  
	We show that the Co atoms confined within corrals on Ag(111) have oscillating
	zero bias anomaly width as a function of corral radius with period on the order of the Ag(111) surface state Fermi wavelength $\lambda_{E_\mathrm{F}}$ \cite{Li2018a-corrals}. 
	We further contrast these results with those of a pure $S=1/2$ Kondo system, 
	providing evidence for the tunability of the \Pc molecule Kondo resonance on Ag(111).
	\begin{figure}[t!]
		\centering
		\includegraphics[width=\columnwidth]{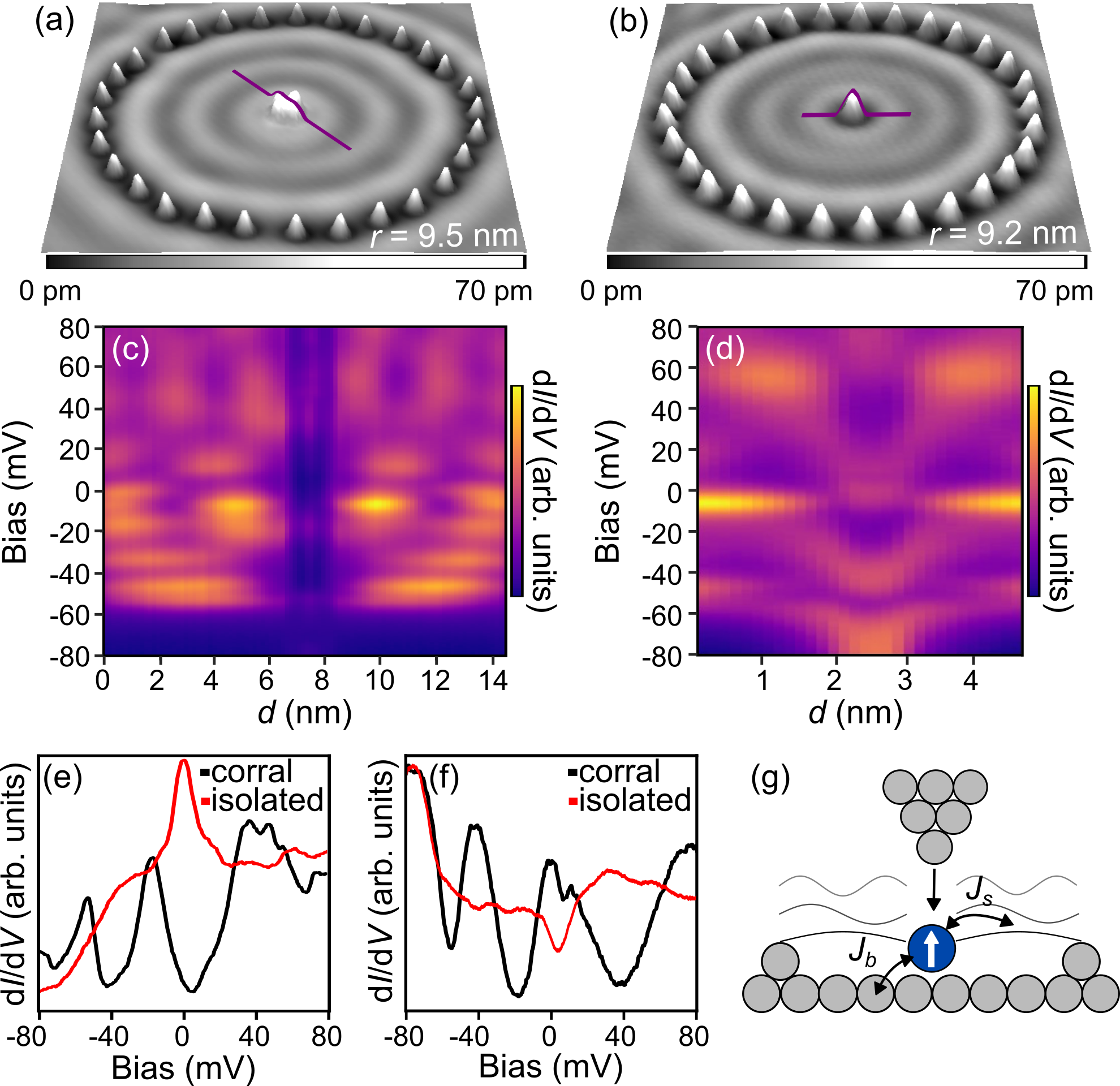}
		\caption{\textbf{(a-b)} STM constant current topography maps of corrals built from Ag adatoms on Ag(111), with a central (a) \Pc molecule and (b) Co adatom ($I= 1$ nA, $V= 80$ mV). \textbf{(c-d)} \dIdV spectra taken across purple lines in (a-b) ($I=1 $ nA, $V_b=80$ mV, $V_\mathrm{mod}=1$ mV). Spectra in (c) normalized by the \dIdV signal at -80 mV. Contrast adjusted for visibility. \textbf{(e)} \dIdV spectra on an isolated \Pc molecule (red) and a \Pc molecule within the quantum corral in (a) (black). \textbf{(f)} \dIdV spectra on an isolated Co adatom (red) and a Co adatom within a $10.1$ nm radius quantum corral (black). \textbf{(g)} Diagram of the measurement where the magnetic impurity interacts with both the bulk and confined surface states of the substrate.}
		\label{fig:fig1}
	\end{figure}
	
	\section{Experimental}
	
	Our experiments were carried out on an Ag(111) crystal (MaTecK GmbH) cleaned by Ne sputtering (voltage $1$ kV, pressure $5\cdot10^{-5}$ mbar) and annealing to 650 \textcelsius ~in UHV ($p< 10^{-9}$ mbar). Atom manipulation and STS was performed at $5$ K in two Createc LT-STM/AFM systems equipped with Createc DSP electronics and control software (version 4.4). Co atoms were evaporated from a thoroughly degassed Co wire wrapped around a W filament and deposited directly onto the Ag(111) sample at $5$ K. \Pc molecules (Sigma Aldrich) were evaporated from a K-cell evaporator at 300 \textcelsius ~onto the substrate at -130 \textcelsius.

	\begin{figure*}[t!]
		\centering
		\includegraphics[width=\textwidth]{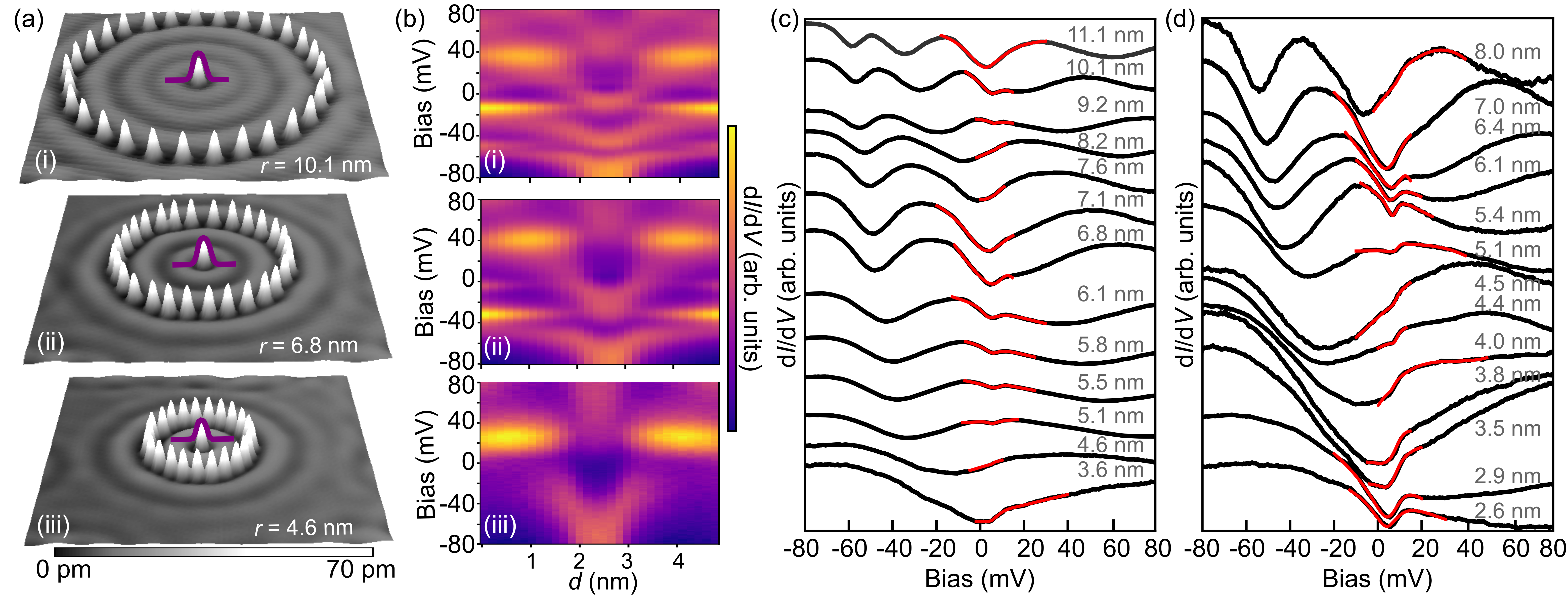}
		\caption{\textbf{(a)} Constant current topography of Co adatoms inside quantum corrals of radius \textit{r} built from Ag adatoms. \textbf{(b)} \dIdV line spectra taken across Co adatoms in Ag corrals with radius $r$ along purple lines in (a). Spectra in subpanels i, ii, iii were measured on the same Co atom, with no STM tip changes from manipulation procedures ($I=1$ nA, $V_b = 80$ mV, $V_\mathrm{mod} = 1-2$ mV
			). \textbf{(c-d)} Representative \dIdV point spectra on Co adatoms in (c) Ag and (d) Co corrals with Fano fits overlaid in red. Fit bounds were selected so $\epsilon_0\approx7.1$ mV to match the value obtained for an isolated Co adatom \cite{Moro-Lagares2019-Co_Ag111-chains, Li2018a-corrals} (see SM Figure S6 \cite{supp} 
			for interpolated spectra and fit values showing $\epsilon_0$ as a function of \textit{r}). Spectra vertically offset for clarity. Typical measurement parameters: $I = 0.5-1.5$ nA, $V_b = 40-1000$ mV, z-offset 0-1 Å.}
		\label{fig:fig2}
	\end{figure*}
	
	After preparing the magnetic impurities and confirming their Fano 
	resonance signal with \dIdV spectroscopy, we proceeded to build the corrals from either Co or Ag adatoms. Individual Ag adatoms were dropped from the tip 
	\cite{Limot2005_PRL_atomTransfer, Kugel2017-atom-dropping} and adatoms were manipulated laterally in constant current mode ($V=2$ mV, $I=60$ nA).
	Figure \ref{fig:fig1}a and b show two complete corrals, where the confined surface state electrons generate a characteristic bias-dependent standing wave pattern in the STM constant-current topography imaging mode.  
	Low bias \dIdV spectra (-80 to 80 mV) were acquired using lock-in amplification with $V_\mathrm{mod}<2$ mV after stabilizing the tip height at 80 mV. Figure \ref{fig:fig1}c and d show \dIdV spectra measured across the corral center, where the confined modes appear as peaks symmetrically around the corral center. The central magnetic impurity acts as a scattering potential for the surface state electrons, modifying the background signal of the confined modes. Figure \ref{fig:fig1}e and f demonstrate how the low energy spectra change within corrals compared to isolated impurities: the zero bias anomaly is not only superimposed on the confined mode background, but is expected to change in shape. Figure \ref{fig:fig1}g illustrates the electronic environment of the magnetic impurity in a quantum corral. The density of states (DOS) at $E_\mathrm{F}$ is composed of bulk conduction bands and the surface states modulated by the corral confinement. Our measurements aimed to extract the Fano resonance signature and determine its evolution as a function of corral radius.

	The low-energy conductance spectra of Co adatoms and \Pc molecules have typically been modelled with a Fano lineshape \cite{Fano}. Observed in tunneling spectra of various Kondo systems, the Fano lineshape arises from interference between electron tunneling paths into a discrete level on the magnetic impurity and a continuum of electronic states in the metallic substrate.
	From a practical perspective, the parameters of a Fano fit can be used to characterize a zero bias anomaly, regardless of its Kondo or spin excitation origin.
	The resulting lineshape in \dIdV spectra is given by 
	\begin{equation}\label{eqn:Fano_res}
		dI/dV(\epsilon)=A\frac{(q+\zeta)^2}{1+\zeta^2} + B\epsilon + C\text{, where }\zeta=\frac{\epsilon-\epsilon_0}{\Gamma_0/2}, 
	\end{equation} 
	$\epsilon$ is energy centered at $E_\mathrm{F}$, $\epsilon_0$ is the energy of the resonant level, $\Gamma_0$ is the resonance full-width at half-maximum, and $A$, $B$, and $C$ are resonance amplitude and linear background coefficients, respectively. 
	STS of isolated Co adatoms (\Pc molecules) on large Ag(111) terraces exhibit a Fano resonance lineshape with $\epsilon_0=8$ meV ($-2$ meV), $\Gamma_0=8$ meV ($11.2 \pm 1.7$ meV) and $q=0.1$ ($\sim 20$) 
	\cite{Li2018a-corrals, Granet2020}. Low-energy tunneling spectra can also vary slightly based 
	on STM tip shape and distance to other surface adatoms and step edges \cite{Moro-Lagares2019-Co_Ag111-chains, Granet2020}. 
	
	To remove effects of varying surface state background \cite{Moro-Lagares2019-Co_Ag111-chains}, we 
	kept the central impurity position constant and only manipulated the corral wall atoms to adjust the surface state density at the center of the corral. We first built the largest corral with the free wall atoms available on the surface, then measured point and line spectra on the central impurity. Then we used semi-automated atom manipulation scripts to measure the corral radius and create a corral slightly smaller than the current one (see Supplementary Material (SM) \cite{supp}). Excess wall atoms were removed as necessary to prevent dimer formation and to keep the mean distance between adjacent wall adatoms 
	smaller than $\lambda_{E_\mathrm{F}}/2$ to control the scattering and confinement of surface state electrons via the corral walls. We built corrals at least 20 nm away from Ag(111) step edges to maintain surface state onset energy at the Co adatom near the nominal value of $-67$ mV (see SM Figure S1\cite{supp}) \cite{Li2018a-corrals,Schneider2002-phaseshift,Li2018-greens-function-approach-corrals}. By building corrals with radii between $2.8-11$ nm, the number of occupied corral eigenmodes was tuned from 0 to 17.

	\clearpage
	
	\section{Results and discussion}
	
	\begin{figure}[t!]
		\centering
		\includegraphics[width=\columnwidth]{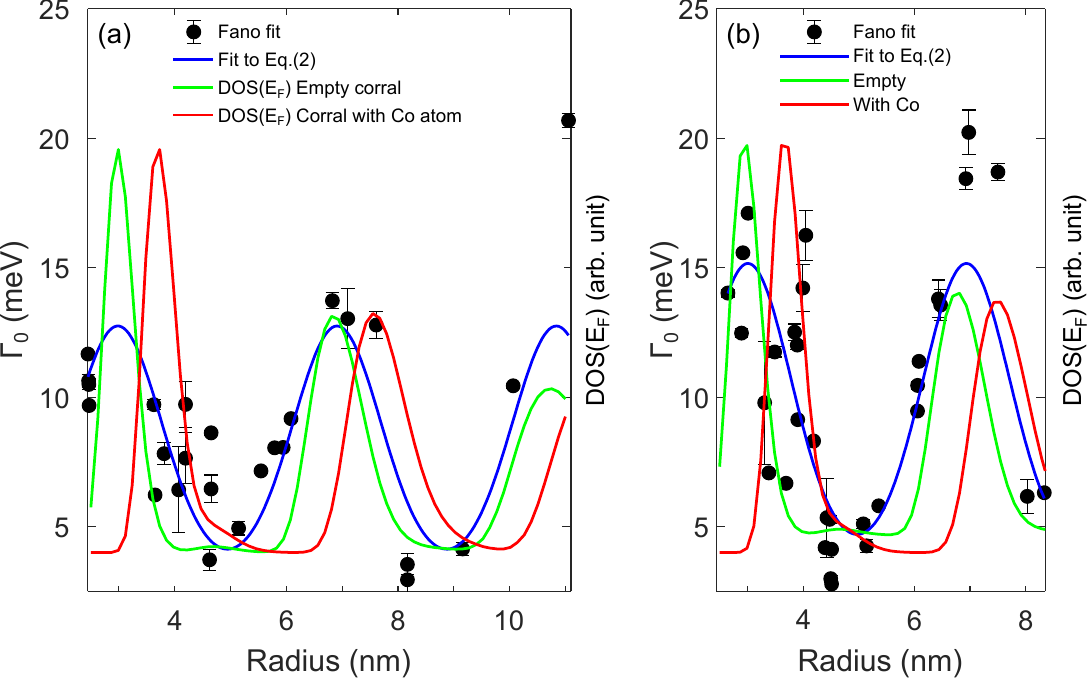}
		\caption{Co adatom Fano resonance width as a function of corral radius for corrals with \textbf{(a)} Ag and \textbf{(b)} Co walls. Fits to Equation~\ref{eqn:model} are overlaid in blue and particle-in-a-box model estimates for LDOS at $E_\mathrm{F}$ in red and green for occupied and empty corrals, respectively. Several spectra were measured on the same atom with different tips and multiple corrals were built with similar radii. Error bars are estimated standard deviations from fit to \ref{eqn:Fano_res}. \change{}{The central Co adatoms were positioned less than a lattice constant away from the fitted corral center point (see SM Figure S12).}}  
		\label{fig:fig3}
	\end{figure}
	\subsection{Co adatoms in Co and Ag corrals}
	Figure \ref{fig:fig2}a shows examples of the built corrals, line spectra measured across the central Co adatom and point spectra on the Co adatom as a function of corral radius. With sufficiently small lock-in modulations and large current setpoints, the zero bias anomaly is seen as a discontinuity close to $\sim$7.1 mV in most corrals. 
	The spatial and energetic profile of the Co adatom bound state near $-80$ meV changes as a function of corral radius (Figure~\ref{fig:fig2}b) as the Co adatom bound state partially overlaps with corral eigenmodes. This signifies modifications in hybridization with the surface band structure \cite{Limot2005-Ag-Co-Ag111-adatom-bound-state-analysis}. The amplitude of the Fano resonance decays approximately 0.5 nm from the center of the Co atom regardless of the corral size, consistent with theory \cite{Plihal2001} and previous experiments for Co/Cu(100) \cite{Knorr2002} and Co/Ag(111) \cite{Moro-Lagares2018-eF_mapping_Kondo}. Similar spatial decay is observed by fitting Fano lineshapes to spectra measured in a grid around a Co atom at the quantum corral center (see SM Figure S4 \cite{supp}).
	
	We fit the \dIdV spectra on the Co atoms in the corral centers with Fano lineshapes and extract the widths $\Gamma_0$ as a function of corral radius. The results plotted in Figure \ref{fig:fig3} display periodic variations in $\Gamma_0$. Earlier studies have associated these variations with the confined modes varying the LDOS at $E_\mathrm{F}$, but the models used to estimate the LDOS variations have been based on empty corral structures \cite{Crommie1993, Kliewer2000a}. 
	The central atom is expected to "gate" the confined modes of a quantum corral \cite{Moon2008-single-atom-gating}, leading to increased confinement and shifting the eigenmode energies localized in the corral center. To assess the impact of this increased confinement, we performed particle-in-a-box calculations for empty corrals and ones occupied with a Co atom approximated as a disk potential \cite{supp}. The parameters needed to initialize the model were obtained from fits to line spectra across corral centers with a Co atom. The general shape of the variation is reproduced by both the empty and occupied corral calculations, but the Fano width data follows the empty corral estimate better.
	This can be rationalised in terms of the Ag(111) surface state dispersion and the resulting Fermi wavelength $\lambda_\mathrm{F} \approx 2\pi/k_\mathrm{F} \approx 75$ Å, which is significantly larger than the Co atom. This means the confined modes of an empty corral are an adequate approximation for the conduction bath environment of the Co atom. In any case, the significant changes of the Fano width can be understood as a consequence of 
	DOS modulation at $E_\mathrm{F}$ due to the quantum confined modes in the corrals. Confined modes on the Co adatom can also be modelled by modifying the surface state dispersion parameters and corral radius (see SM Figures S2 and S3 \cite{supp}). 
	
	In order to compare with previous result in the literature \cite{Li2018a-corrals}, we now estimate surface and bulk state exchange constants between Co and Ag(111) by using the model that incorporates the width $\Gamma_0$ as a function of corral radius $r$: 
	\begin{equation}
		\label{eqn:model}
		\Gamma_0 = D\exp \left(\frac{\change{}{-}1}{J_b\rho_b+ J_s\rho_{s0}(1+A\cos(2kr+\delta))}\right).
	\end{equation}

	We fix $\rho_{s0}=0.125$ eV$^{-1}$, $\rho_b=0.27$ eV$^{-1}$, $D=4.48$ eV in accordance with \cite{Li2018a-corrals} and fit equation \ref{eqn:model} to the extracted Fano widths $\Gamma_0$ for Co adatoms in 
	corrals with Ag and Co walls. From the fit we extract exchange coupling energies $J_{s,b}$ and scattering phase shift $\delta$. Several different fit models accurately reproduce the measured trend in the zero bias anomaly width as a function of corral radius. Furthermore, some models reproduce the trend well but have large covariance between variables. We treat these models separately in the SM \cite{supp}. Fit results for the simplest model with $A=1$ are compiled in Table~\ref{table:1} and visualized in Figure~\ref{fig:fig3}.
	
	\begin{figure*}[t!]
		
		\includegraphics[width=\textwidth]{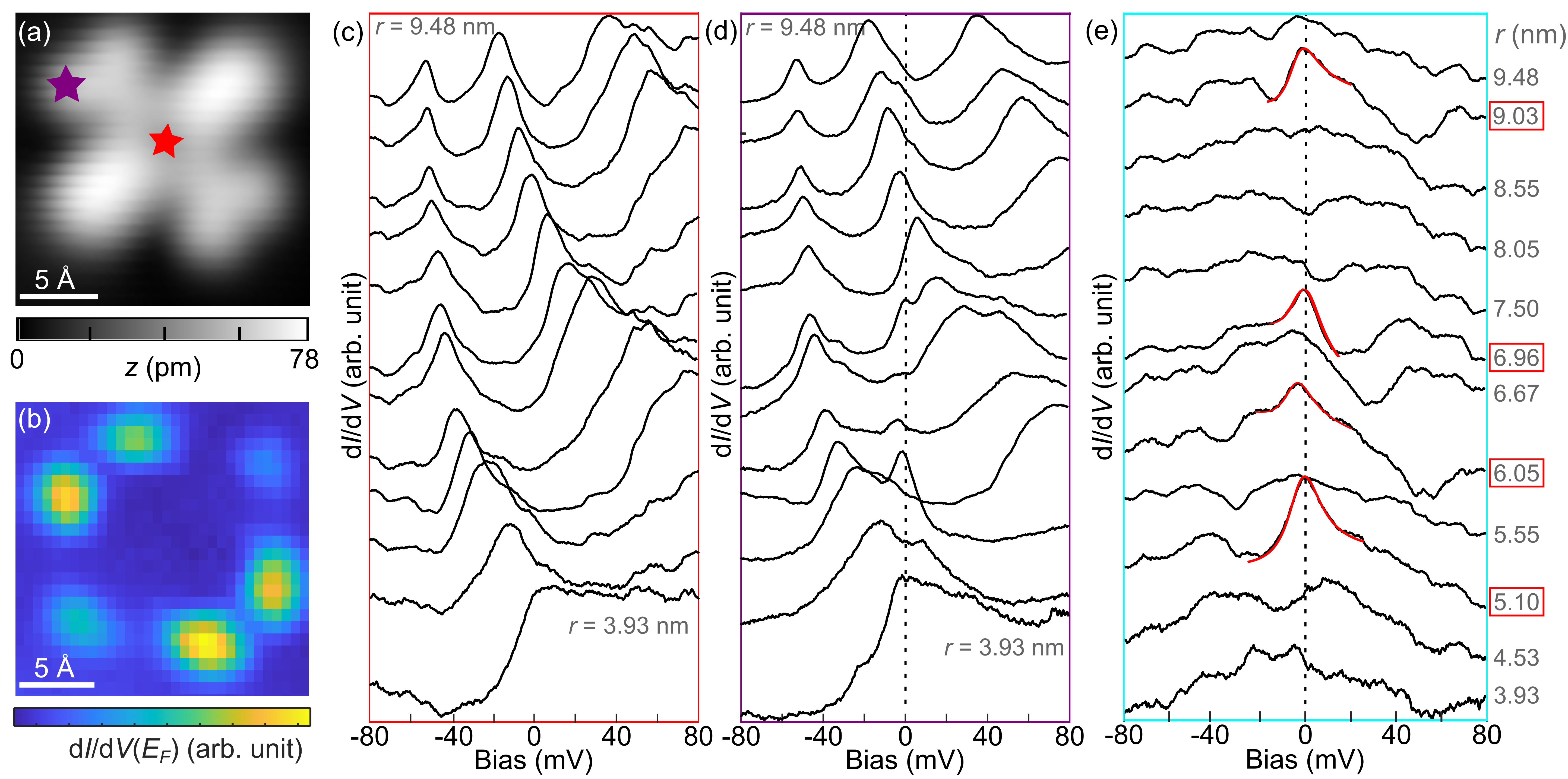}
		
		\caption{\textbf{(a)} Constant current topography of an isolated \Pc molecule ($I = 1$ nA, $V = 80$ mV). \textbf{(b)} Zero bias conductance map of an isolated \Pc molecule. \textbf{(c)} \dIdV spectra taken at the center of \Pc molecules at the center of Ag adatom quantum corrals of radius $r$. \textbf{(d)} \dIdV spectra taken at the split lobe of \Pc molecules at the center of Ag quantum corrals of radius $r$. \textbf{(e)} \dIdV spectra from (d) with \dIdV spectra from (c) subtracted, in order to better detect the Kondo resonance peak. Fano resonance fits overlaid in red. 
		}
		\label{fig:fig4}
	\end{figure*}

	\begin{table}[ht]
		\centering
		\caption{Model fit parameters to Co/Ag(111) data}
		\label{table:1}
		\begin{tabular}{ c|c|c } 
			\hline
			Parameter & Ag corrals & Co corrals \\
			\hline
			$J_s$ (eV) & \change{0.120$\pm$0.019}{0.11$\pm$0.02} & \change{0.122 $\pm$0.018}{0.12 $\pm$0.02}\\ 
			$J_b$ (eV) & 0.530$\pm$0.013 & \change{0.536$\pm$0.013}{0.54$\pm$0.015} \\ 
			$\delta$ & \change{1.515 $\pm$ 0.127}{1.511 $\pm$ 0.132} & \change{1.600$\pm$0.141}{1.47$\pm$0.14} \\
			\hline
		\end{tabular}
	\end{table}
	
	Extracted exchange constants $J_{b,s}$ for Co/Ag(111) are consistent with those reported by Li et al.~\cite{Li2018a-corrals} for corrals built from both Co and Ag adatoms. We see variations in Fano resonance energy $\epsilon_0$ and asymmetry parameter $q$ as a function of corral radius (see SM Figure S6 \cite{supp}). 
	For small corral radius $r$, choice of corral wall adatom may affect the lifetime or spin polarization of surface state electrons coupled to the central impurity. Limot et al.~report that Ag adatoms alter Ag(111) surface state lifetime whereas Co adatoms do not \cite{Limot2005-Ag-Co-Ag111-adatom-bound-state-analysis}. 
	Surface state quasiparticle lifetimes impacted by the wall atom species may affect the amplitude of resonance features seen on the central Co adatom \cite{Fernandez2016}. 
	Moro-Lagares et al.~showed that the Fano resonance amplitude, and none of the other resonance parameters, 
	depends on the type of adatom in dimers and coupled chains on Ag(111); only Co-Ag and Co-Co dimers within a distance less than six lattice sites showed any differences in Fano resonance amplitude \cite{Moro-Lagares2019-Co_Ag111-chains}. For all our corrals, distances between the central and wall atoms were larger than distances where an effect on the Fano resonance would be noticeable. Summarizing, as seen in Table~\ref{table:1}, we only observe the expected change in the scattering phase shift $\delta$ but no difference in either $J_s$ or $J_b$ between Ag and Co corrals.

	The data presented here can be used for comparison with the recently proposed spinaron origin of the observed zero-bias anomaly. The observed width variation is significantly stronger than expected if the features arose from simple spin-flip excitations of a higher spin magnetic impurity with non-zero magnetic anisotropy energy. The spinaron excitation, in contrast, is expected to be sensitive to the density of states of the corral due to the induced polarization in the bath.
	The spinaron model of magnetic impurities on Cu, Ag and Au(111) surfaces suggests that the lineshape should fit the Fano resonance (rather than e.g.~the Fano-Frota lineshape \cite{Frota}) and that the resonance should be shifted towards positive bias by several mV in contrast to the classic Kondo resonance that is very close to $E_\mathrm{F}$ \cite{Bouaziz2020-spinaron, friedrich2023spinaron,Noei2023}. Interestingly, these criteria are fulfilled in the Co/Ag(111) system, yet predictions for the detailed variation of the resonance width with changing the surface state DOS would require a full first principles many-body treatment of this system. 
	
	\subsection{\Pc molecules in Ag corrals}
	
	To establish whether the Kondo resonance of other magnetic impurities can be tuned via quantum confinement, we repeated the measurements with \Pc molecules inside corrals built from Ag adatoms.
	As shown in Figure \ref{fig:fig4}b, the Kondo resonance of an isolated \Pc molecule is localized on the frontier orbitals which hold the impurity spin. Local variations in Fano fit parameters of an isolated molecule were established from grid spectroscopy (see SM Figure S5 \cite{supp}).

	For the \Pc molecules inside the corrals, \dIdV spectra were measured along the 'split-lobe' direction across the molecule to characterize both the Kondo resonance and the confined modes (examples of spectra along a line across the molecule are shown in SM Figure S7 \cite{supp}). Figure~\ref{fig:fig4}c and d show the \dIdV spectra measured on the molecule center (red star in Figure \ref{fig:fig4}a) and on the split lobe (purple star in Figure \ref{fig:fig4}a) for varying corral radii. 
	The expected Kondo peaks are not clearly distinguishable in most of the spectra on the split lobe due to the confined mode background and its spatial variations around $E_\mathrm{F}$. In particular, there are modes with a node in the corral center that are not visible in the spectra shown in Figure \ref{fig:fig4}c, but contribute to the features in Figure \ref{fig:fig4}d (see SM Figure S8 \cite{supp}). Heat map plots of the line spectra show contrast close to $E_\mathrm{F}$ for many corrals, suggesting that the Kondo effect persists even if it is smeared by the confined modes (see SM Figure S8 \cite{supp}). The \dIdV curves in Figure \ref{fig:fig4}d can be fitted well with 4-8 Gaussians, but significant Kondo peaks do not stand out by comparing the fit results with particle-in-a-box calculations (see SM Figure S9 \cite{supp}).

	Spectra at the molecule center were subtracted from the split lobe point spectra in order to minimize background DOS effects.
	Figure \ref{fig:fig4}e shows the result of this subtraction: at certain radii, a peak at zero bias becomes visible. Fano lineshapes were fitted to low bias peaks in the subtracted point spectra. For radii $r=6.05$ and $6.96$ nm, the Kondo resonance does not overlap with the 
	confined modes. For corral radii $r = 5.1$ and $r = 9.03$ nm, there are some contributions from local variations of confined modes across the molecule length as indicated by particle-in-a-box modelling \cite{supp}; however, considering the spectral weight of the other confined modes remaining after the subtraction, we can still extract the Fano linewidth with sufficient confidence. For the rest of the corrals, quantitative extraction of the Fano parameters was not possible. 
	Extracted fit parameters are shown in Table \ref{tab:PcFanoParams}.
	
	\begin{table}[h]
		\caption{Fano fit parameters for \Pc molecules.}\label{tab:PcFanoParams}
		\begin{tabular}{c|c|c|c}
			\hline
			\Pc environment & $\Gamma_0$ (meV) & $\epsilon_0$ (meV) & q (arb. unit)  \\ \hline
			5.1 nm radius corral & 17.1 & -2.0 & 5.5 \\
			6.05 nm radius corral & 16.1 & -3.9 & 4.5 \\
			6.96 nm radius corral & 15.3 & 0.5 & 32.8 \\
			9.03 nm radius corral & 15.6 & -4.8 & 2.0 \\
			Isolated & 25.5 & -2.4 & 7.2 \\ \hline 
		\end{tabular}
		
	\end{table}
	
	Significant variations are visible in the width, energy position and the $q$-factor across the different corral radii. The widths extracted are consistently smaller than those on isolated molecules, as with previous measurements on self-assembled monolayers \cite{Granet2020}. Although the lack of visible Kondo peaks on most radii makes a fit to equation \ref{eqn:model} impossible and thus the exchange constants cannot be extracted, the corral has a clear impact on the Kondo features as compared to an isolated molecule. 
	
	The overall lineshape and behaviour are strikingly different compared to Co adatoms inside circular corrals, where the Fano dip is clearly distinguishable for most corral radii. The spin state, impurity orbital energies, and orbital overlaps with the conduction bands are markedly different between Co atoms and \Pc molecules, which should result in a different amplitude for the Fano width variation according to equation \ref{eqn:model}. In addition, it is likely that the zero-bias features correspond to spinaron excitation for the cobalt atom, and a clean $S=1/2$ Kondo effect for the \Pc molecules.  
	A number of experimental factors could influence the observed width and precise shape of the \Pc Kondo peak, such as adsorption geometry \cite{Granet2020}, tip DOS \cite{Moro-Lagares2018-eF_mapping_Kondo} and instrument noise conditions \cite{Gruber2018}. We measured the Kondo signatures of an isolated \Pc molecule as a function of tip height and find only limited variation in the resulting Fano fit parameters (see SM Figures S10 and S11 \cite{supp}). Therefore variations in the tip height or the tip-molecule interaction do not explain the different radius dependence and suppressed Fano signatures compared to the Co atoms.
	
	To probe the influence of confined modes on the \Pc Kondo resonance further, an accurate model of the confined mode energy widths and lifetimes on the molecule is required. 
	Recent advances towards electron spin resonance STM on Ag surfaces \cite{Esat2023} hold promise for exploring the magnetic properties of \Pc molecules in quantum corrals. Given the complexity uncovered in this system, further experiments with other Kondo impurities (such as triangulenes \cite{Turco2023}) in quantum corrals and theoretical studies on their behaviour may also be warranted. 
	
	\section{Conclusions}
	We have shown the tunability of spinaron and Kondo excitations by probing individual magnetic impurities in atomically-engineered quantum corrals.
	Our findings demonstrate 
	the significance of the surface state in the low-energy spinaron and Kondo zero bias anomalies of magnetic impurities adsorbed on Ag(111). We observed periodic variations in the zero bias anomaly widths of Co/Ag(111) by modulating the surrounding surface state and extracted exchange constants consistent with those previously reported. These results confirm that the Co adatom zero bias anomaly is coupled to the DOS at $E_\mathrm{F}$ and demonstrate qualitatively different behaviour between the spinaron Co atom and Kondo \Pc molecules in the electronic environment generated by quantum corrals. Our findings establish a starting point to control spinaron and Kondo excitations via confinement engineering of an underlying conduction bath.
	
	\section*{Acknowledgements} 
	We thank G. Chen, R. Drost and X. Huang for assistance with the preliminary experiments and fruitful discussions. This research made use of the Aalto Nanomicroscopy Center (Aalto NMC) facilities and Aalto Research Software Engineering services. The authors acknowledge funding from the European Research Council (ERC-2017-AdG no.~788185 ``Artificial Designer Materials''and ERC-2021-StG no.~101039500 ``Tailoring Quantum Matter on the Flatland'') and the Academy of Finland (Academy professor funding nos. 318995 and 320555 and Academy research fellow nos.~331342, 336243 and nos.~338478 and 346654). 
	
	M.A.~and A.K.~contributed equally to this work.
	
	\bibliography{refs}
	
\end{document}


\renewcommand\thefigure{S\arabic{figure}}
\renewcommand\thetable{S\arabic{table}}

\newcommand{\change}[2]{#2}

\title{Supplemental Material for `Tuning spinaron and Kondo resonances via quantum confinement'}

\author{Markus Aapro}
\email[Corresponding authors. ]{markus.aapro@aalto.fi, abraham.kipnis@aalto.fi, peter.liljeroth@aalto.fi}
\affiliation{Department of Applied Physics, Aalto University, Finland}

\author{Abraham Kipnis}
\email[Corresponding authors. ]{markus.aapro@aalto.fi, abraham.kipnis@aalto.fi, peter.liljeroth@aalto.fi}
\affiliation{Department of Neuroscience and Biomedical Engineering, Aalto University, Finland}

\author{Jose L. Lado}
\affiliation{Department of Applied Physics, Aalto University, Finland}

\author{Shawulienu Kezilebieke}
\affiliation{Department of Physics, Department of Chemistry and Nanoscience Center, 
University of Jyväskyl\"a, FI-40014 University of Jyväskyl\"a, Finland}

\author{Peter Liljeroth}
\email[Corresponding authors. ]{markus.aapro@aalto.fi, abraham.kipnis@aalto.fi, peter.liljeroth@aalto.fi}
\affiliation{Department of Applied Physics, Aalto University, Finland}

\maketitle
\newcommand{\Pc}{H$\mathrm{_2}$Pc~}
\newcommand{\dIdV}{d$I$/d$V$~}
\newcommand{\MA}[1]{\textcolor{blue}{{\bf MA: #1}}}

\section{Ag(111) surface state dispersion}
\begin{figure}[ht]
    \centering
    \includegraphics{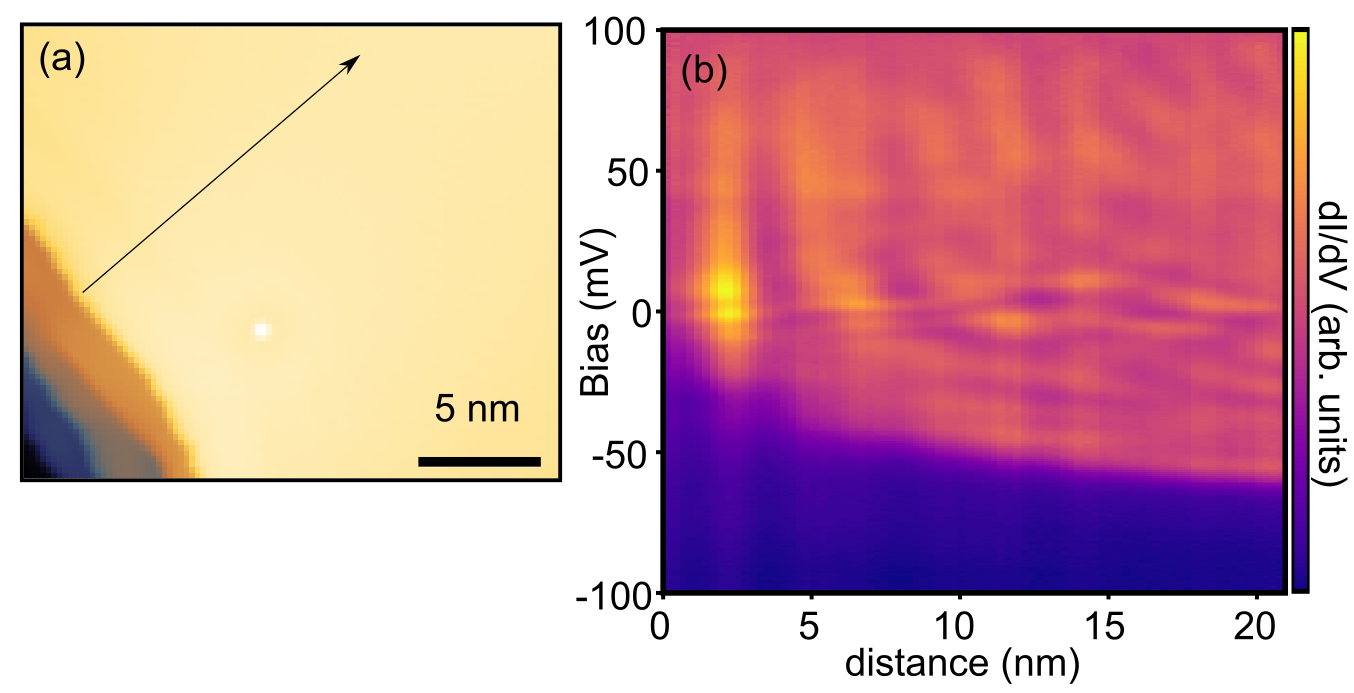}
    \caption{\textbf{(a)} Constant current topography of clean Ag(111) step edge mostly free of defects. Bias 100 mV, current 500 pA. \textbf{(b)} \dIdV spectra as a function of distance from Ag(111) step edge, across the line marked in (a). Current 500 pA, lock-in modulation amplitude 2 mV. Ag(111) surface state electron phase coherence 
    length and lifetime can be estimated from these data \cite{Wahl2003, Jensen2005}.}
    \label{fig:S1}
\end{figure}

STM tip condition was verified before performing spectroscopy by first measuring a \dIdV spectrum on a terrace of Ag(111) free of defects. A flat spectrum, with the onset of the Ag(111) surface state band at -67 mV, indicates flat tip density of states necessary for the Tersoff-Hamann approximation. If spectroscopy did not show a sharp onset at $E_0=-67$ mV, the tip was reshaped by indentation into the Ag(111) and field emission until smooth spectra were acquired and individual atoms on the surface were imaged as spheres. Corrals were built at a distance farther than 20 nm from a step edge to control for the background in the surface state onset energy. In Figure~\ref{fig:S1}a we show a constant current topography of a 
Ag(111) terrace. In Figure~\ref{fig:S1}b we show \dIdV line spectra taken across the line in Figure~\ref{fig:S1}a to illustrate the change in surface state background near Ag(111) step edges. 

\section{Interpolated \dIdV spectra and analytical particle-in-a-box model}
\begin{figure}[h]
    \centering
    \includegraphics[width=0.8\textwidth]{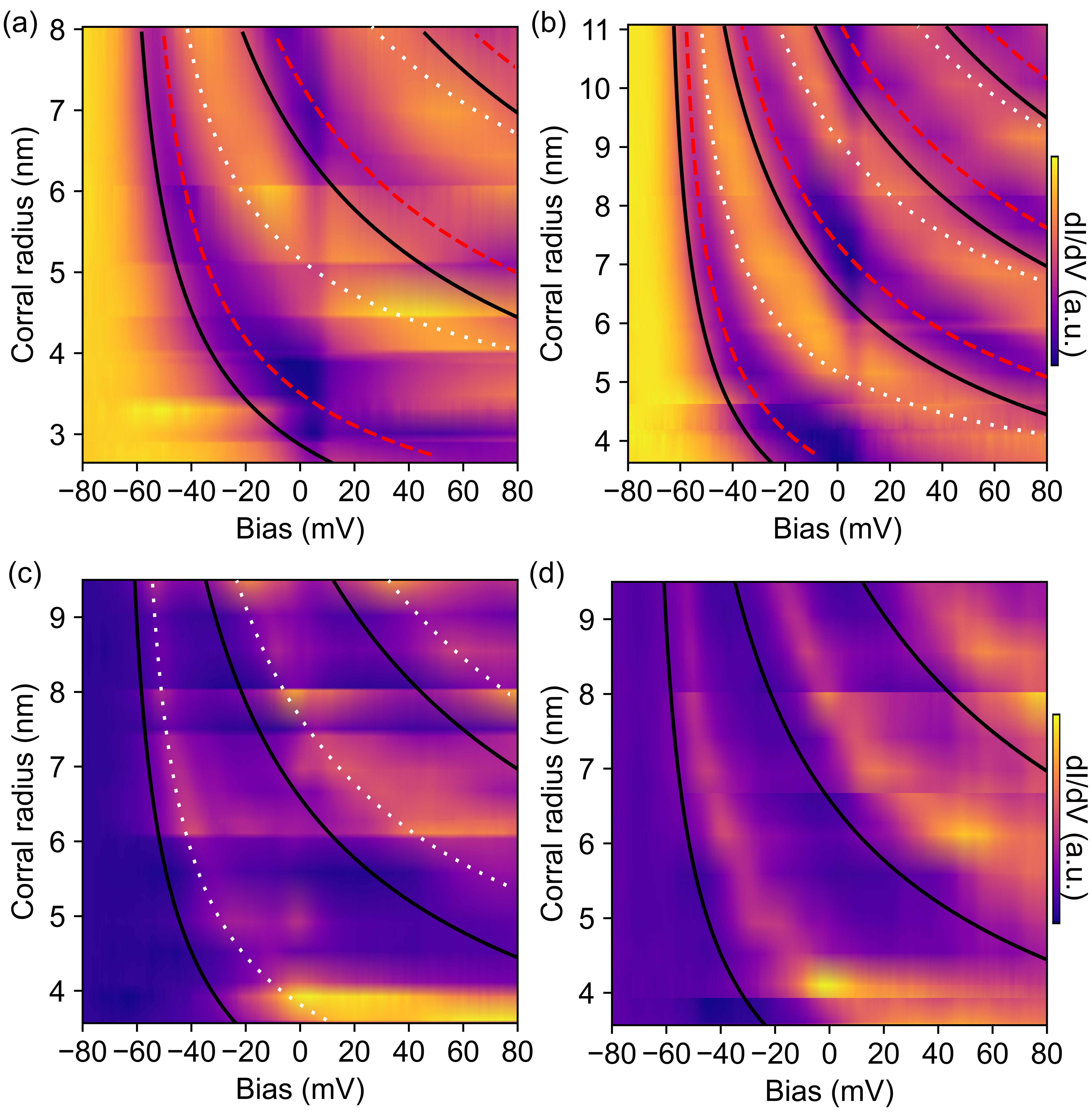}
    \caption{Interpolated \dIdV spectra for corrals with central Co adatom and Ag \textbf{(a)}  and Co \textbf{(b)} walls. \textbf{(c)} Spectra at the edge of \Pc molecules inside corrals made from Ag atoms. \textbf{(d)} Spectra at the center of \Pc molecules inside corrals made from Ag atoms.
    Solid lines: predictions using equation \ref{eqn:hard_wall_model} from hard-walled particle-in-a-box model for eigenenergies of empty corral, using $m^*=0.4m_e$ and $E_0=-67$ mV. Dotted white lines: eigenenergies from equation \ref{eqn:hard_wall_model} with $m^*=0.4m_e$ and $E_0=-55$ mV (a-b), $-60$ mV (c), and radius offset 2 nm (a-b), 0.8 nm (c). Dashed red lines: eigenenergies using $m^*=0.4m_e$, $E_0=-60$ mV, $r_0=0.4$ nm. }
    \label{fig:interpolated_spectra}
\end{figure}

Data in Figure~\ref{fig:interpolated_spectra} show the \dIdV spectra as a function of radius interpolated on a radius mesh with $dr=0.1$ nm. Overlaid in solid lines are solutions of the Schrödinger equation for eigenenergies of a free electron in a hard walled circular box: 
\begin{equation}
    E_n = \frac{\hbar^2}{2m^*}(\frac{J_0(n)}{r})^2-E_0
    \label{eqn:hard_wall_model}
\end{equation} 
where $m^*$ is the effective mass of the surface state electrons in Ag(111) $m^*=0.4m_e$, $E_0=-67$ mV is the Ag(111) surface state onset energy, $J_0(n)$ is the $n$th zero of the Bessel function $J_0$ and $r$ is the corral radius. Eigenenergies of the empty hard-walled corral model overlap the empty energy gaps between eigenmodes in the occupied corrals, similar to results seen in Mn corrals on Ag(111) \cite{Kliewer2001}. Color scales are normalized for each curve by \dIdV at $-80$ mV.

\begin{figure}[H]
    \centering
    \includegraphics[width = .9\textwidth]{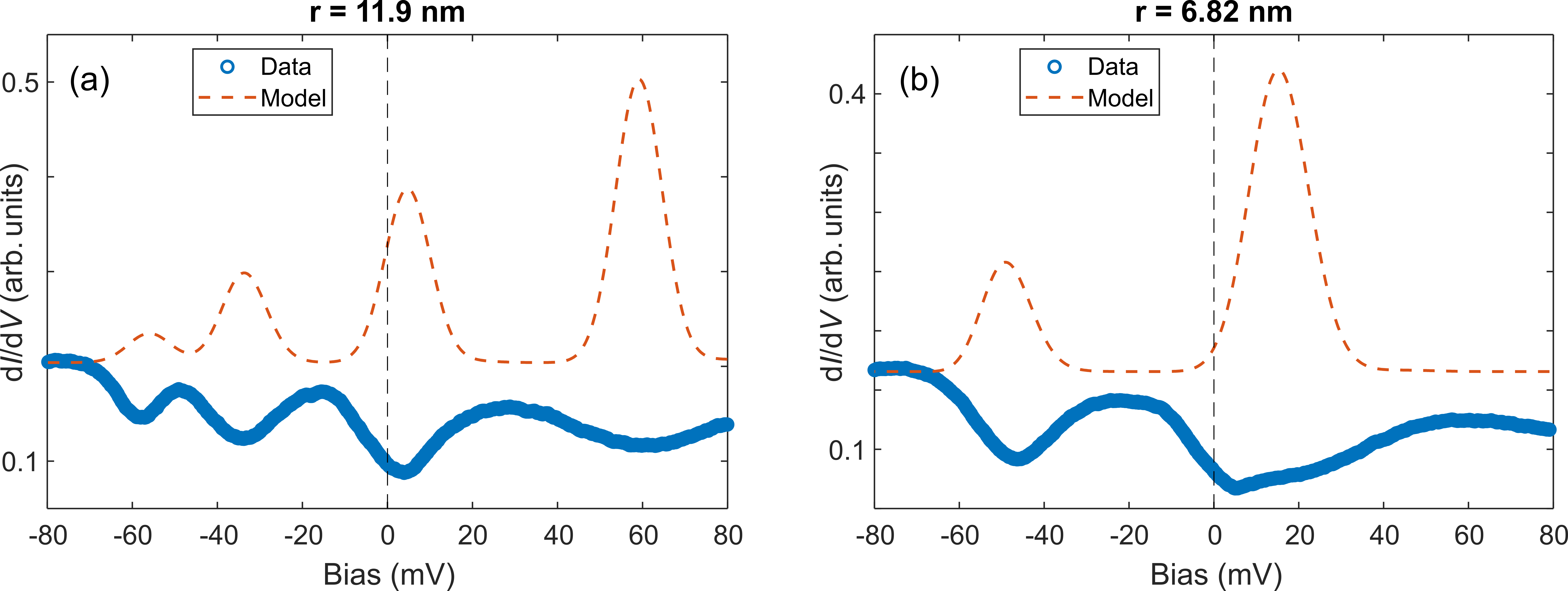}
    \caption{\dIdV spectra on central Co atoms in quantum corrals with Ag walls (blue circles) and corresponding LDOS estimate from a particle-in-a-box model fitted to eigenmodes around the Co atom (red dashed lines). \textbf{(a)} for a corral of radius $11.9$ nm and \textbf{(b)} $6.82$ nm. Dips in the \dIdV spectra align with modelled confined mode peaks with better than $\sim3$ meV accuracy on the bias/energy axis, in line with previous works \cite{Kliewer2000a}.}
    \label{fig:CoCenterSpectraModelling}
\end{figure}

To compare the impact of the central impurity on the confined modes of the corral, we manually fit the eigenmodes returned by Equation~\ref{eqn:hard_wall_model} to the signatures of confined modes in our \dIdV spectra. For the Co adatom, the confined modes correspond to dips in the spectra: this is in line with previous studies \cite{Kliewer2000a} and our numerical particle-in-a-box calculations as shown in Figure \ref{fig:CoCenterSpectraModelling}. For comparison, we also fit the conductance maxima on Co atoms. For the \Pc molecule we interpret the conductance peaks as signatures of confined modes. 
We introduce a `radius offset' $r_0$ and replace $r$ with $r-r_0$ in Equation~\ref{eqn:hard_wall_model}. We also float the value of $E_0$, which may change slightly depending on the confinement of electrons near step edges or inside corrals (see Figure~\ref{fig:S1}). For \Pc corrals we extract $E_0=-55$ mV and $r_0=0.8$ nm and for Co we extract $E_0=-60$ mV and $r_0=0.4$ nm. These values indicate that the \Pc molecule gates the eigenmodes of the corral more than the Co adatom, despite the lesser hybridization with the Ag orbitals seen in the absence of an inversion effect. We note that these values do not fully describe the physical picture, i.e. an extracted value of $E_0$ shifted from the nominal value of -67 mV may be a result of lifetime effects, i.e. a broadening of the surface state band edge due to the impurities, rather than a change in the location of the gaps of the projected band in Ag(111).

\section{Spatial variation of Fano fit parameters}

\begin{figure}[H]
    \centering
    \includegraphics{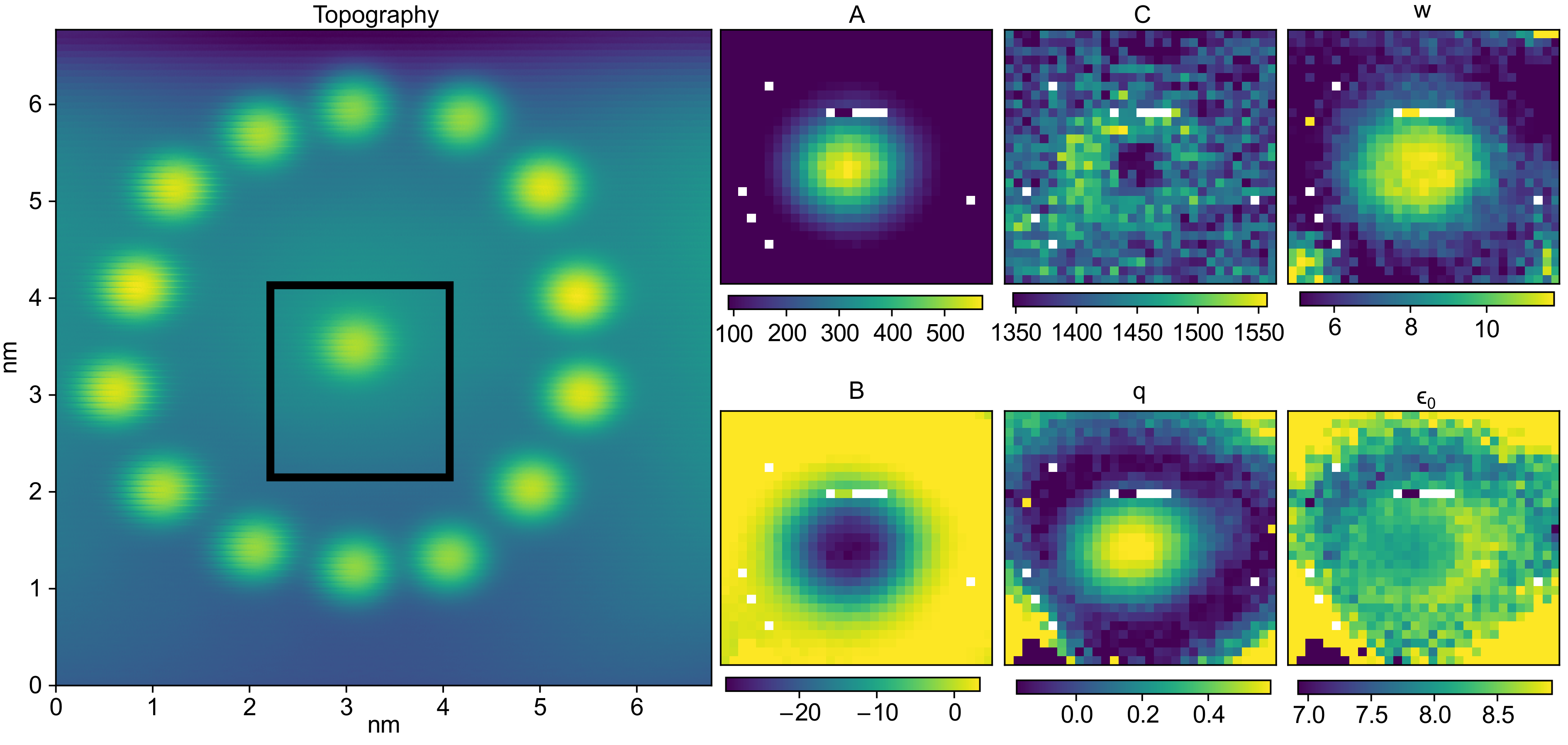}
    \caption{Left: constant current topography of a 2.5 nm radius quantum corral with Ag adatom walls and central Co adatom. Right: Fit parameters extracted from fitting the Fano resonance (equation in the main text) to spectra taken in a grid pattern over the central atom in the black box on the left. White pixels indicate the fit did not converge. All parameters decay over approximately 0.5 nm from the center of the adatom.}
    \label{fig:Co_FanoFits_grid}
\end{figure}

\begin{figure}[H]
    \centering
    \includegraphics[width = \textwidth]{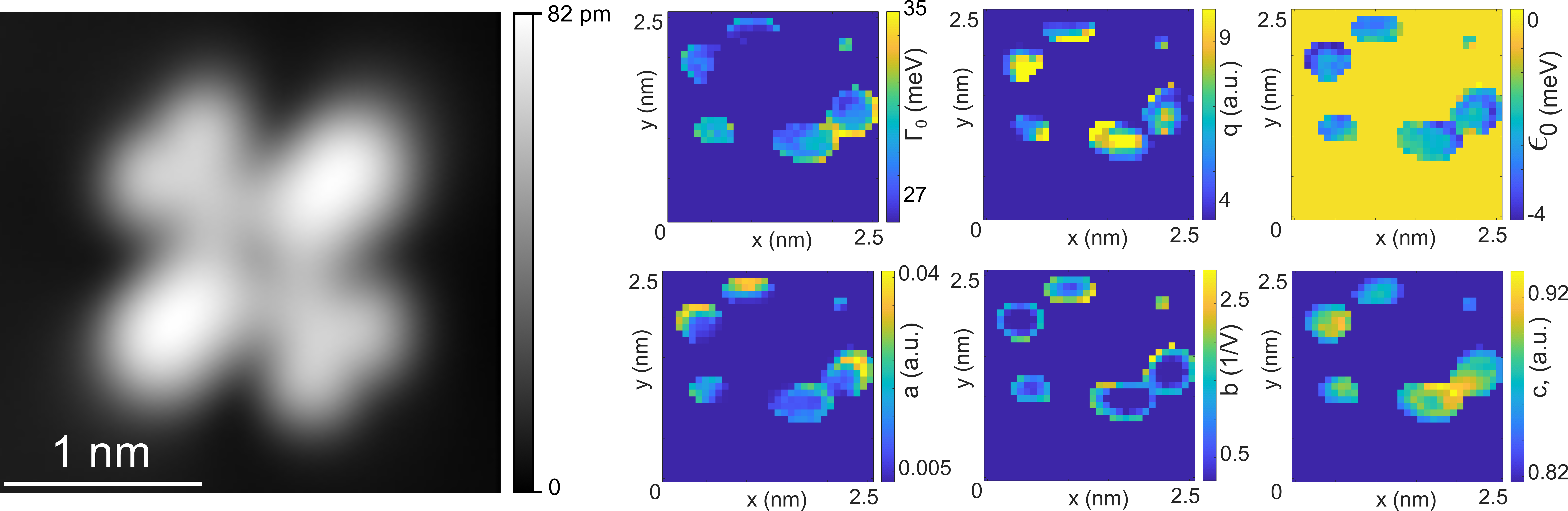}
    \caption{Left: constant current topography of an isolated \Pc molecule ($I = 1$ nA, $V = 80$ mV). Right: Fit parameters extracted from fitting the Fano resonance to \dIdV grid spectra on the \Pc molecule.}
    \label{fig:H2Pc_FanoFits_grid}
\end{figure}

\section{Other fit parameters ($\epsilon$, q) for Co/Ag(111) corrals}

\begin{figure}[h]
    \centering
    \includegraphics[width=\textwidth]{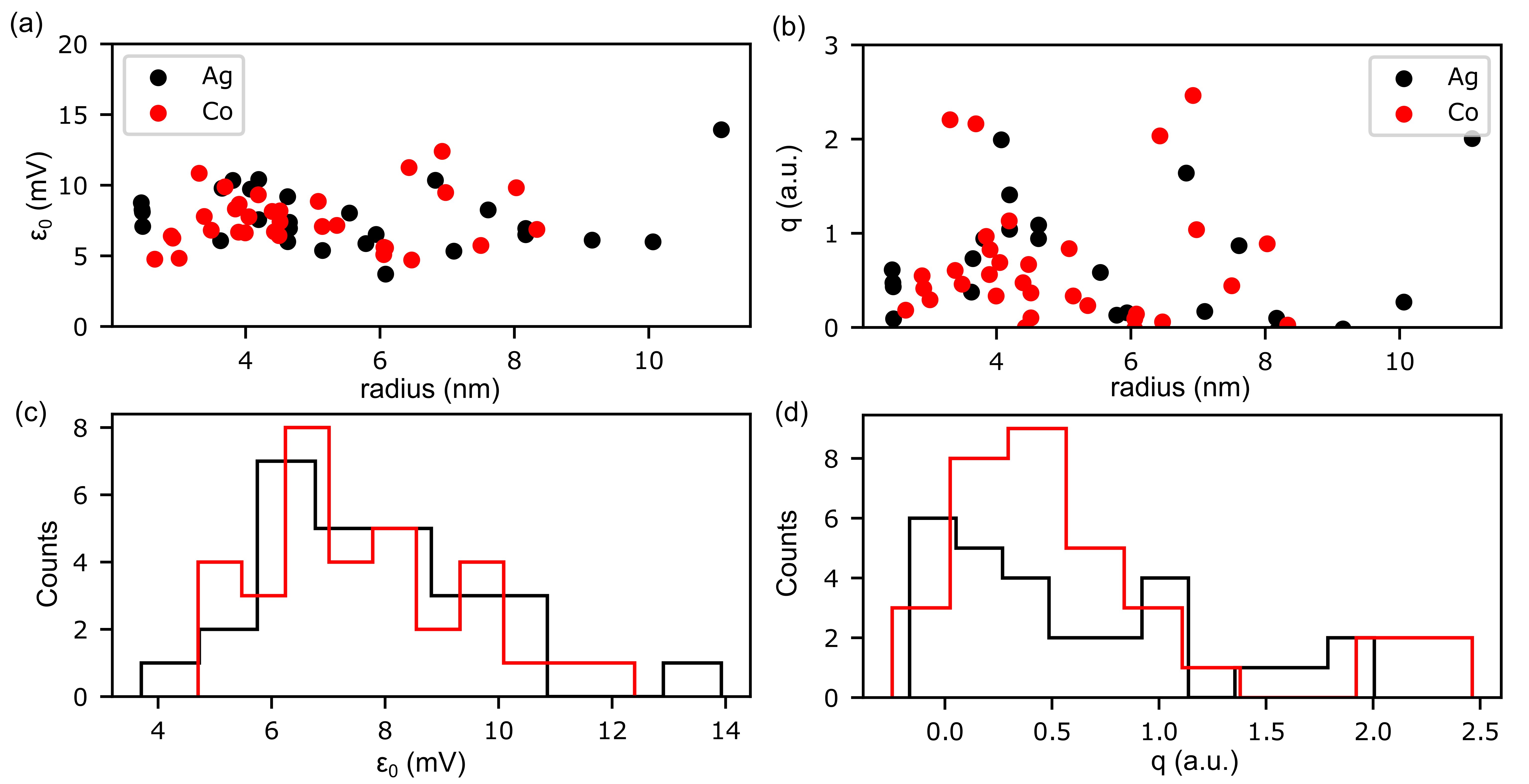}
    \caption{\textbf{(a-b)} Values for $\epsilon_0$ and $q$ from Fano fits on central Co atom for corrals made from Ag and Co walls. \textbf{(c-d)} Histograms of the values in (a-b). The values for $\epsilon_0$ are centered around 7.1 mV and the values for q are centered around 0.4. Our observation of a slight periodic change between asymmetric Fano and Lorentzian lineshape (evident in the asymmetry parameter q of the Fano resonance fit) is consistent with theoretical predictions by Usjaghy et al.~\cite{Ujsaghy2000}.}
    \label{fig:fit_histograms}
\end{figure}

\section{Phenomenological models for fitting couplings $J_s$, $J_b$}
Zero bias anomaly resonance width $w$ follows the dependence 
    $w\propto De^{-1/(J\rho)}$,
where $D$ is the bandwidth of the conduction band coupled to the impurity state
, $J$ is the strength of the coupling between the conduction band and the impurity
, and $\rho$ is the DOS at the impurity site
\cite{Nagaoka_1965}. We separate $J\rho$ into surface DOS and bulk DOS contributions as 
$J\rho=J_s\rho_s + J_b\rho_b$
where $J_s$ is the coupling strength to the surface state electrons, $J_b$ is the coupling strength to bulk electrons, and $\rho_{s,b}$ are the surface and bulk state electronic DOS, respectively. Considering Friedel oscillations from single impurities in a two-dimensional electron gas, we describe the DOS a distance $r$ from impurities as  
\begin{equation}
    \label{eqn:rho_r}
    \rho_s(r) = \rho_{s0}(1+\frac{A\cos(2kr+\delta)}{(kr)^\alpha})
\end{equation} where $\rho_0$ is a constant background surface state density, $A$ is the amplitude of the modulating term, $k$ is the surface state wave vector at $E_\mathrm{F}$
, $\delta$ is the scattering phase shift and $\alpha$ is the decay constant \cite{Crommie1993}. We choose $\alpha=0$ to simplify the model and reduce the risk of overfitting: as shown in figure 3 in the main text, we observe negligible decay in the Fano width variation for the measured corral radii. 
We use the previous equations to form a phenomenological model for
the zero bias anomaly width $w$ as a function of corral radius $r$: 
\begin{equation}
\label{eqn:model}
    w = D\exp \left(\frac{\change{}{-}1}{J_b\rho_b+ J_s \rho_{s0}(1+A\cos(2kr+\delta))}\right).
\end{equation}
We fit our data with different variations of this model to explore the different potential physical interpretations of the data.

\begin{table}[ht]
\centering
\caption{Model fit parameters to Co/Ag(111) data}
\label{table:1}
\begin{tabular}{ c|c|c } 
 \hline
 ID & Equation \\ \hline
 A& $D\exp(-1/(J_b\rho_b+J_s\rho_{s_0}(1+(0.3)*\cos{(2kx+\delta)}) )$\\ 
 B& \change{$D\exp(-1/(J_b\rho_b+J_s\rho_{s_0}(1+A\cos{(2kx+\delta)}) )$}{$D\exp(-1/(J_b\rho_b+J_s\rho_{s_0}(1+\cos{(2kx+\delta)}) )$}\\
 C& \change{$D\exp(-1/(J_b\rho_b+J_s\rho_{s_0}(1+\cos{(2kx+\delta)}) )$}{$D\exp(-1/(J(\rho_b+\rho_{s_0}(1+A\cos{(2kx+\delta)})) )$}\\
 D& \change{$D\exp(-1/(J(\rho_b+\rho_{s_0}(1+A\cos{(2kx+\delta)})) )$}{$D\exp(-1/(J(\rho_b+\rho_{s_0}(1+\change{}{(0.3)*}\cos{(2kx+\delta)})) )$}\\
 \hline
\end{tabular}
\end{table}
\begin{table}[ht]
\centering
\caption{Model fit parameters to Co/Ag(111) data. For all models, $D=4480$ \change{}{meV}, $\rho_{s_0}=0.125$ eV$^{-1}$, $\rho_{b}=0.27$ eV$^{-1}$, $k=0.8$ nm\change{}{$^{-1}$}.}

\begin{tabular}{|l|l|l|l|}
\hline
                            & \textbf{Fit params} & \textbf{Ag}     & \textbf{Co}     
                            \\ \hline
\multirow{3}{*}{\textbf{A}} & $J_s$ (eV)             & \change{$0.40\pm0.06$}{$0.38\pm0.07$}   & \change{$0.40\pm0.06$}{$0.41\pm0.07$}    \\ \cline{2-4} 
                            & $J_b$ (eV)               & \change{$0.40\pm0.03$}{$0.41\pm0.03$}    & \change{$0.40\pm0.03$}{$0.41\pm0.07$}   \\ \cline{2-4} 
                            & $\delta$            & $1.51\pm$\change{$0.12$}{$0.13$}   & \change{$1.60\pm0.14$}{$1.47\pm0.15$}   
                            \\ \hline
\multirow{3}{*}{\change{\textbf{C}}{\textbf{B}}} & $J_s$ (eV)               & \change{$0.12\pm0.02$}{$0.11\pm0.02$}   & $0.12\pm0.02$   \\ \cline{2-4} 
                            & $J_b$ (eV)               & $0.53\pm0.01$   & \change{$0.53\pm0.01$}{$0.54\pm0.02$}   \\ \cline{2-4} 
                            & $\delta$            & \change{$1.5\pm0.12$}{$1.51\pm0.13$}    & \change{$1.6\pm0.14$}{$1.47\pm0.15$}    \\ \hline
\multirow{3}{*}{\change{\textbf{D}}{\textbf{C}}} & $J$ (eV)                   & \change{$0.402\pm0.004$}{$0.40\pm0.004$}   & \change{$0.400\pm0.004$}{$0.408\pm0.005$}   \\ \cline{2-4} 
                            & $A$                   & $0.29\pm0.05$   & \change{$0.30\pm0.04$}{$0.30\pm0.05$}   \\ \cline{2-4} 
                            & $\delta$            & \change{$1.5\pm0.12$}{$1.51\pm0.13$}    & \change{$1.6\pm0.14$}{$1.47\pm0.15$}    \\ \hline
\multirow{2}{*}{\change{\textbf{E}}{\textbf{D}}} & $J$ (eV)                   & \change{$0.353\pm0.006$}{$0.4\pm0.003$} &  \change{$0.348\pm0.005$}{$0.4\pm0.003$}\\ \cline{2-4} 
                            & $\delta$            & \change{$1.45\pm0.10$}{$1.5\pm0.12$}   & \change{$1.32\pm0.11$}{$1.47\pm0.14$}   \\ \hline
\end{tabular}
\end{table}
All models effectively capture the observed trends in the data. Caution is warranted as some models exhibit overfitting, with high cross-correlation between certain variables, i.e.~$J_{\rho, b}$ and $A$. Notably, all models consistently show insignificant differences between the data for corrals constructed from Co versus Ag walls.

\section{Numerical particle-in-a-box model for confined modes}

We use a particle-in-a-box model to solve the 2D Schrödinger equation in circular geometries: 

\begin{equation}
    \centering
    \hat{H} \psi_n = \left(-\frac{\hbar^2}{2m^{*}}\nabla^2 + V(\vec{r}) \right)\psi_n  = \left(E_n + E_0\right) \psi_n
\end{equation}

Here $m^{*}$ is the effective mass and $E_0$ is the onset energy of the surface state band. The Co atoms and H$_\mathrm{2}$Pc molecules inside the corral are modeled as regions with constant potential $V(\vec{r})$. After initialising the necessary surface state dispersion parameters, atom/molecule potentials and geometry, the eigenmodes $\psi_n$ and their associated eigenenergies $E_n$ are calculated with the finite element method in Matlab. 
\begin{figure}[!t]
    \centering
    \includegraphics[width = 0.89\textwidth]{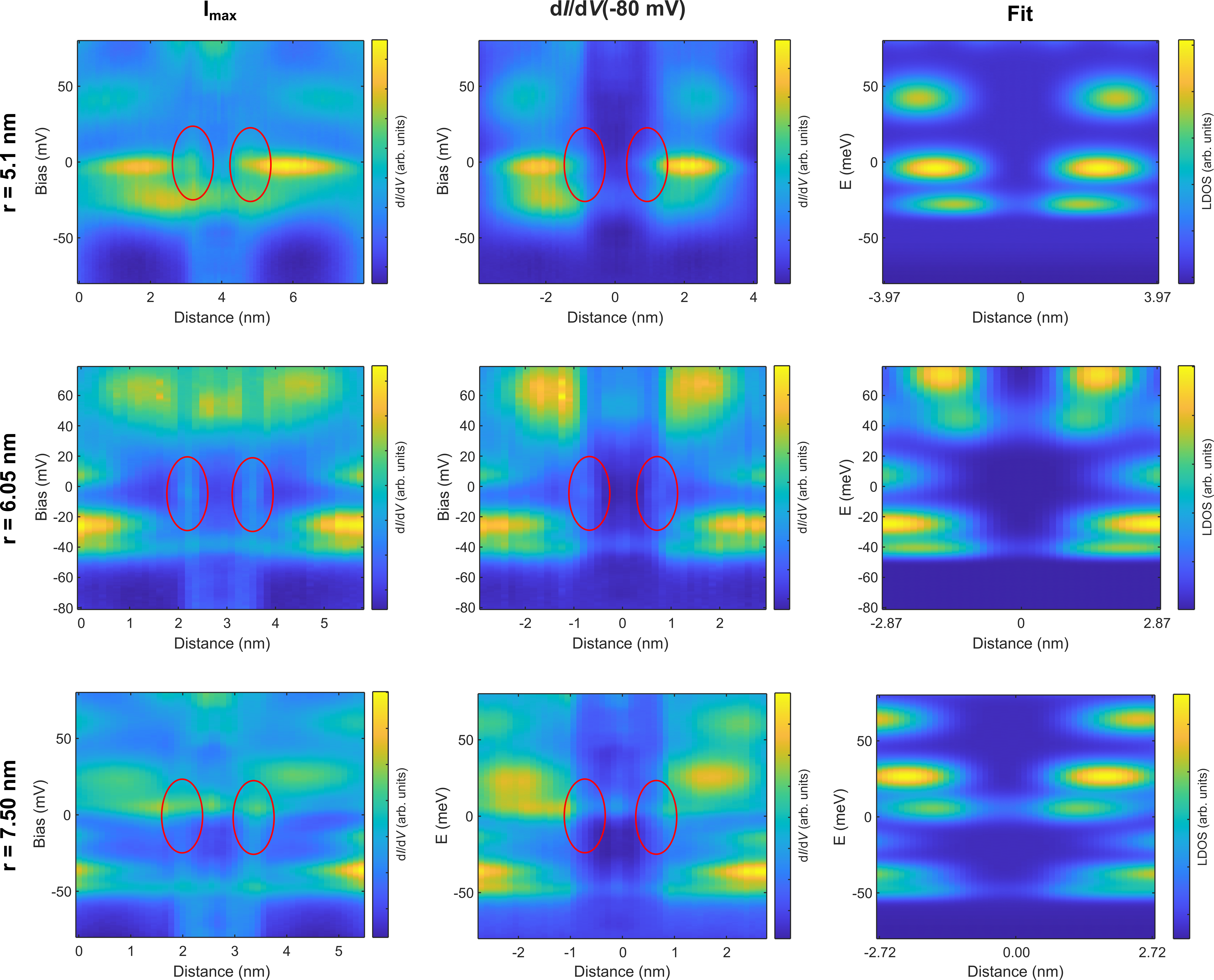}
    \caption{Typical examples of line spectra with two normalisation methods and eigenmode fit results. Columns from left to right: line spectra normalised with maximum current ($I_{max}$) at each point, line spectra normalised to \dIdV at $-80$ mV and the corresponding fit to the data. Rows from top to bottom: data for corral radii $5.1$ nm, $6.05$ nm and $7.50$ nm. The particle-in-a-box model was fitted to the \dIdV data normalised to $-80$ mV. Red ovals highlight the features close to $E_\mathrm{F}$ on the split lobe of the molecule.}
    \label{fig:SI_LS_fit_examples}
\end{figure}
To fit the line spectra in corrals of various sizes and wall atom spacings, the LDOS associated with the eigenmodes is modelled using Gaussians with energy-dependent widths. This approach has been found effective in quantum corrals built from CO molecules on Cu(111) \cite{Weiss2023Broadening}, where the confined mode linewidths follow a simple relation

\begin{equation}
    \centering
    \Gamma_W(E) = \frac{\hbar}{d}\sqrt{\frac{2(E - E_0)}{m^{*}}},
\end{equation}

where $d$ is the average distance between wall atoms. We find that our confined mode widths do not follow this relation directly, possibly due to scattering from the zero bias anomaly in the center and/or differences in lifetime effects between Cu(111) and Ag(111). Hence we use a linear correction and express the Gaussian widths as

\begin{equation}
    \centering
    \Gamma(E) = D\Gamma_W(E) + E_B
\end{equation}

where $D$ and $E_B$ are constants used as fitting parameters. Finally, a broadened step function at $E_0$ (approximating the surface state spectrum outside the corral) is added to account for the imperfect confinement in the corrals. Figure \ref{fig:SI_LS_fit_examples} shows example fits to \dIdV line spectra across a corral with a \Pc molecule in the center.

\section{Modelling LDOS at $E_F$ with and without a central Co atom}

The particle-in-a-box model was used to estimate variations of the LDOS at $E_F$ in the corral center, which is expected to influence the observed the zero bias anomaly widths on Co adatoms. To this end, we fitted line spectra data for several corral radii with the particle-in-a-box model using the following fit parameters:

\begin{itemize}
    \item Corral radius
    \item Eigenmode broadening $E_B$
    \item Atom potential
    \item Surface state step height
    \item Eigenmode decay coefficient $D$
    \item Surface state onset energy $E_0$
    \item Lateral offset along the major axis
\end{itemize}

The Co atom was approximated with a disk potential of radius $r = 0.55$ nm. The averaged fit parameters were used to calculate the LDOS at $E_F$ at the center of a corral with and without a central Co atom, shown in the main Figure 3. 
We observe that the empty corral eigenmodes serve as a good approximation to the observed zero bias anomaly width variation. 

\clearpage

\section{Confined modes and Fano peaks in \Pc spectra}

\begin{figure}[H]
    \centering
    \includegraphics[width = \textwidth]{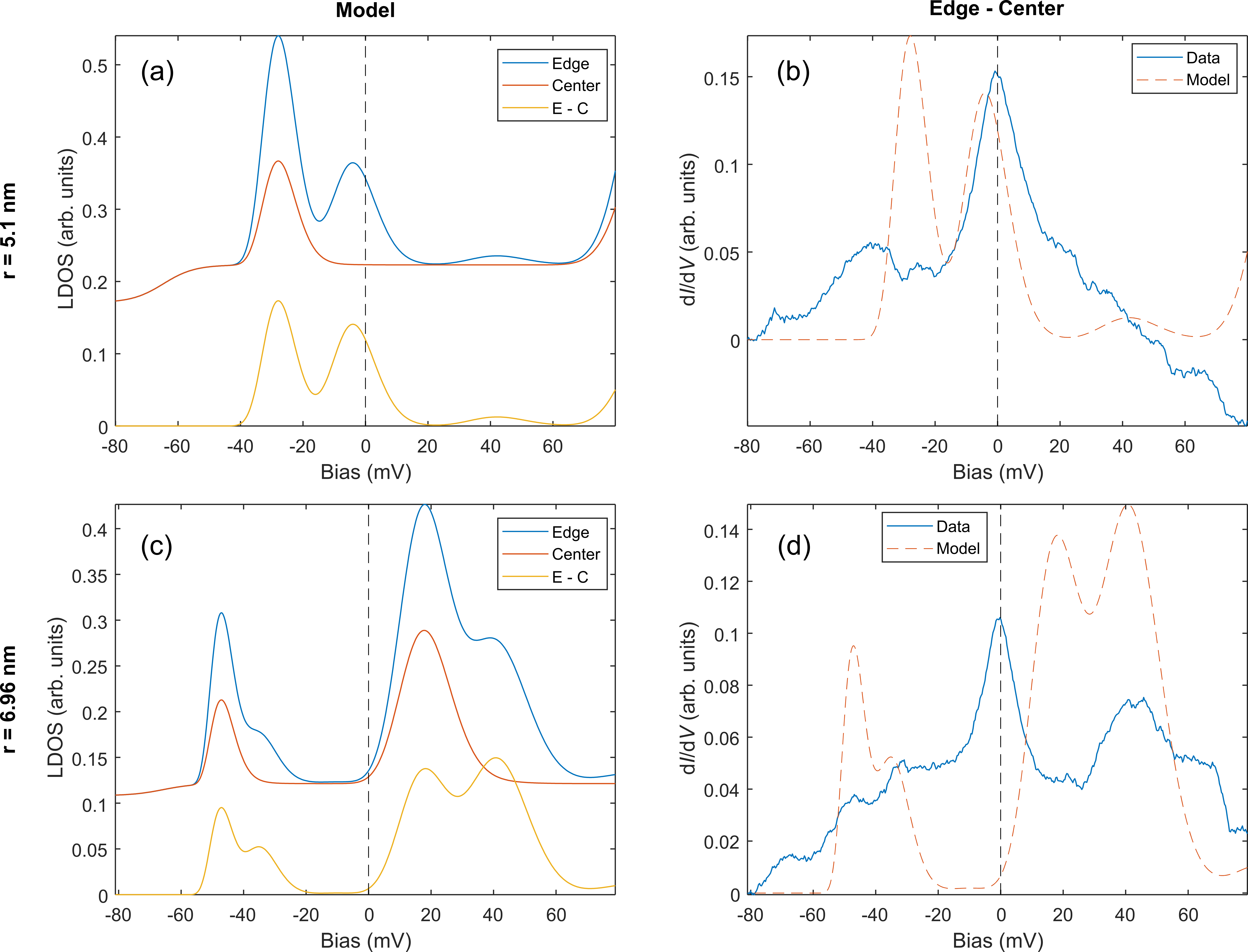}
    \caption{Contributions of confined modes to observed zero bias conductance on \Pc molecules. \textbf{(a)} Modelled confined mode spectra at the edge (blue) and center (red) of a \Pc molecule, and their difference (yellow). Corral radius $r = 5.1$ nm. \textbf{(b)} The measured (solid blue line) and modelled (dashed red line) difference spectra. The zero bias peak coincides with expected contrast from confined modes, although the relative intensities are not normalised. \textbf{(c-d)} Corresponding plots for corral radius $r = 6.96$ nm. The zero bias peak does not coincide with confined modes.}
    \label{fig:SI_Pc_E-C_model}
\end{figure}

\clearpage
\section{Gaussian sum fits to the \Pc corral spectra}

\begin{figure}[H]
    \centering
    \includegraphics[width = \textwidth]{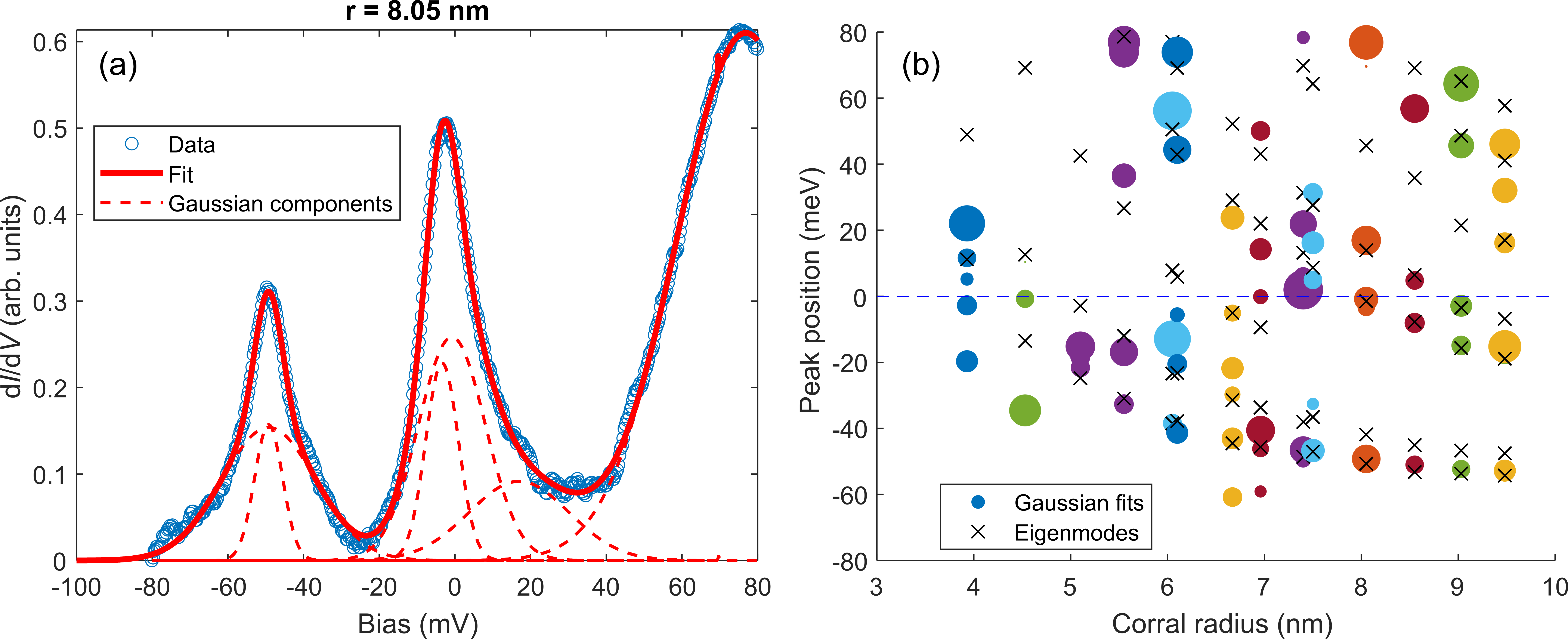}
    \caption{Gaussian fits to the \dIdV spectra measured on the split lobes of \Pc molecules inside quantum corrals. \textbf{(a)} an example spectrum with a corresponding Gaussian sum fit. \textbf{(b)} the energy positions of the fitted Gaussians (circles) and eigenmode energies (crosses) as a function of corral radius. Circle radius signifies the width of the fitted Gaussian. Only eigenmodes with a non-zero amplitude around the corral center are plotted.} 
    \label{fig:gauss}
\end{figure}

\clearpage

\section{Tip height variation in STS of \Pc molecules}

\begin{figure}[H]
    \centering
    \includegraphics[width = .85\textwidth]{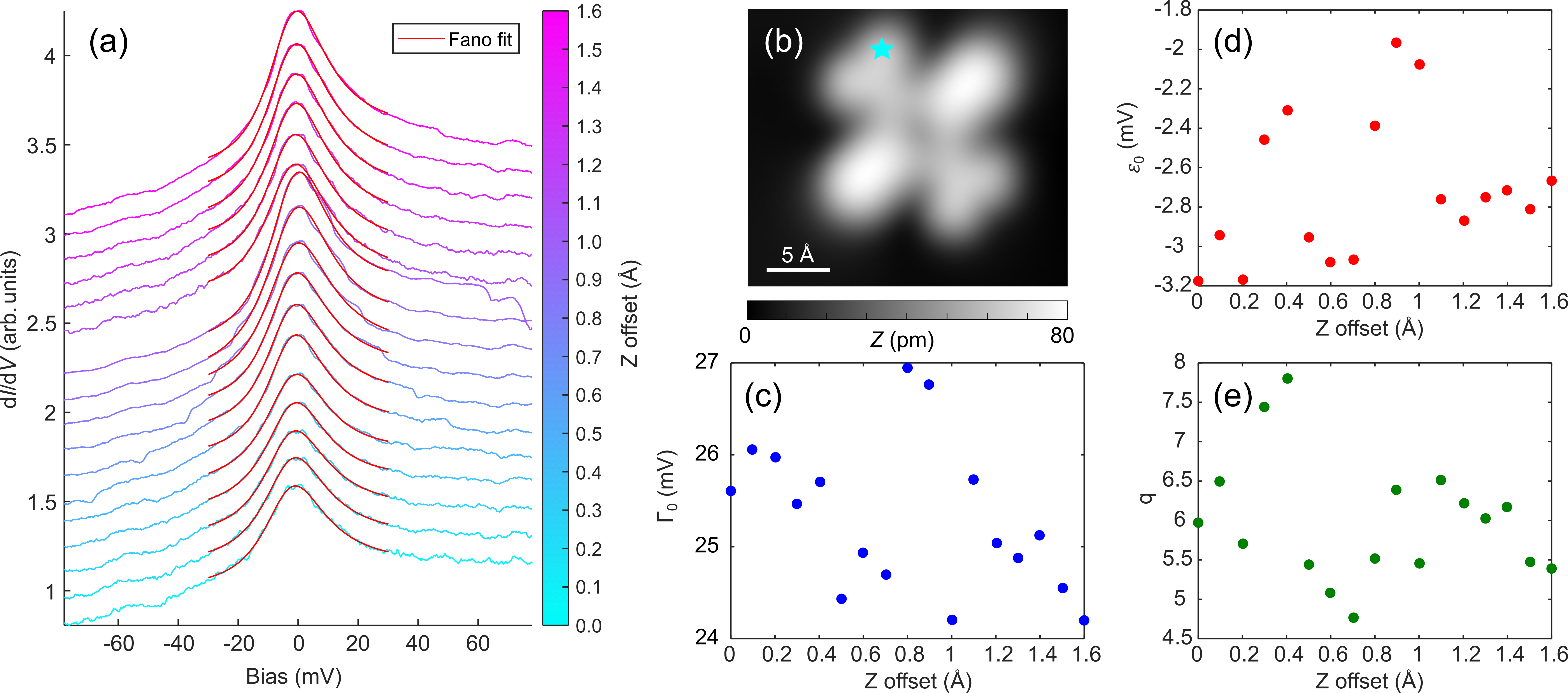}
    \caption{\textbf{(a)} Point spectra on an isolated \Pc molecule (the same one as in main Figure 4a). Tip height stabilised at a setpoint of $1$ nA and $80$ mV before opening feedback and applying the z-offset towards the surface. Resulting currents at $80$ mV range from $1$ to $30$ nA. Fano lineshapes were fitted in an energy range of $\pm 30$ mV. The steps in the spectrum corresponding to z-offset 1 Å are due to preamplifier current saturation. \textbf{(b)} The \Pc molecule and spectrum position (light blue star). Scan setpoint $1$ nA, $80$ mV. \textbf{(c-e)} Fano fit parameters of the curves shown in (a). The fit results show limited variation and possibly a very faint downward trend for the Fano width.}
    \label{fig:Pc_heightDependence}
\end{figure}

\begin{figure}[H]
    \centering
    \includegraphics[width = 0.6\textwidth]{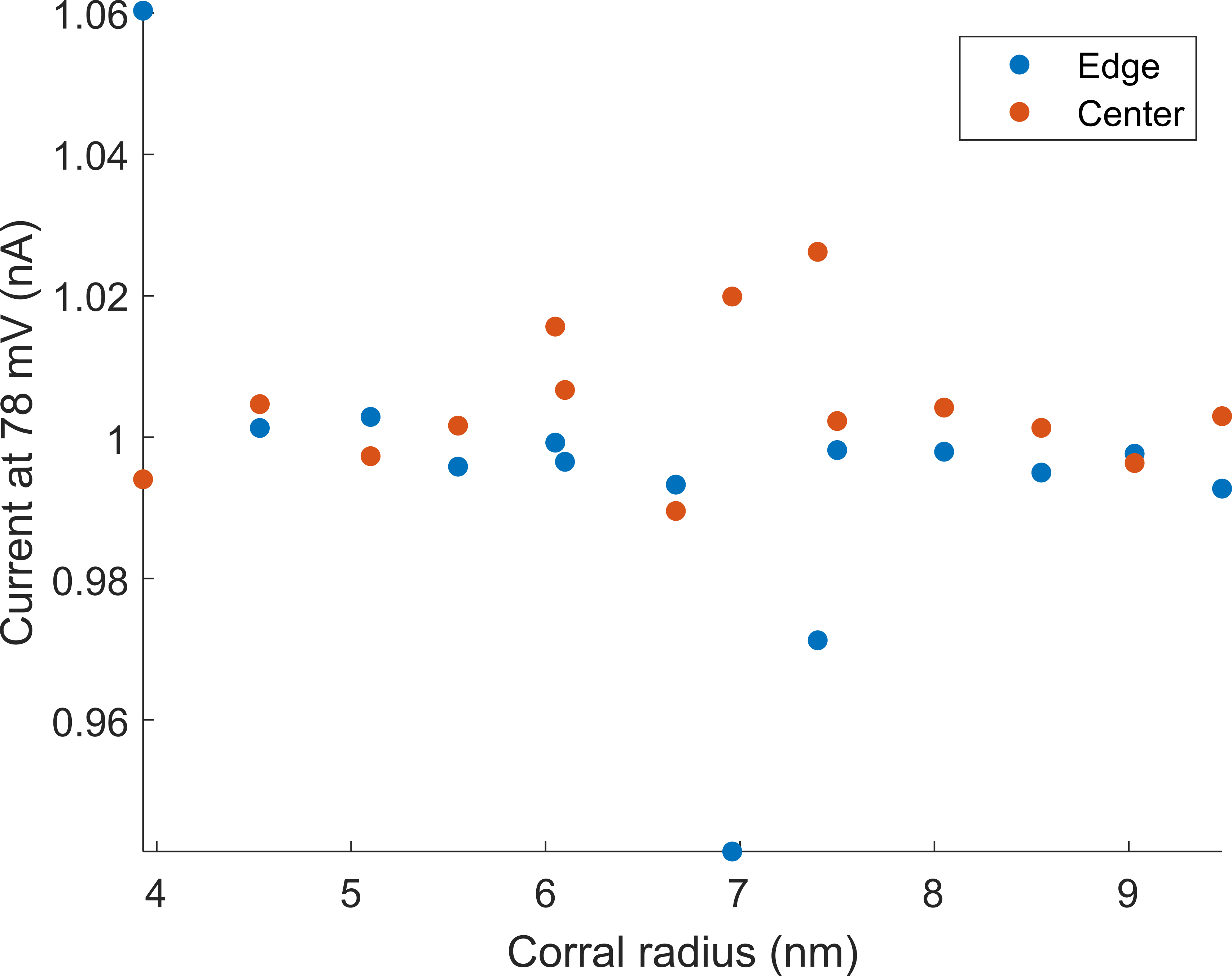}
    \caption{Tunneling currents at $V = 78$ mV in the STS measurements shown in main Figure 4. The current varies up to $50$ pA from the average value due to slightly different tip height stabilisation methods and changes in conductance spectra.}
    \label{fig:Pc_StartCurrents}
\end{figure}

\change{}{\section{Co adatom distances from corral centers}
\begin{figure}[h]
    \centering
    \includegraphics[width=0.8\textwidth]{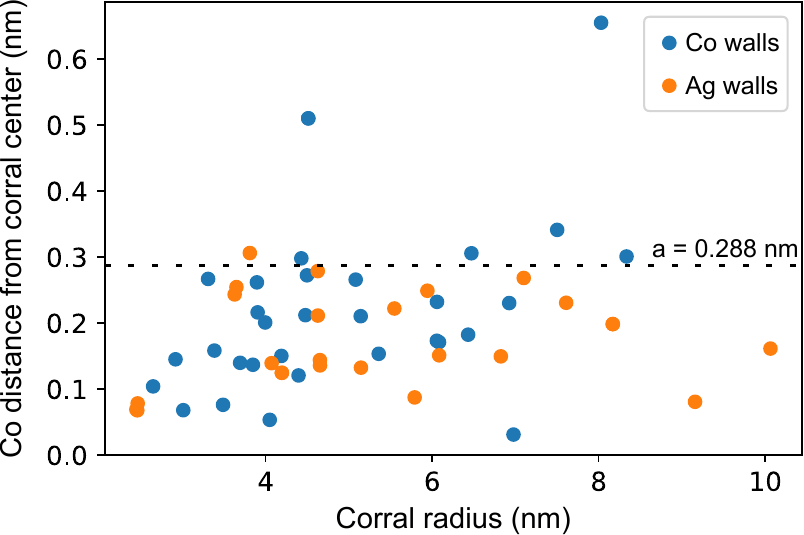}
    \caption{Co atom distances from the fitted circular corral centers. For most corrals the distance is less than a Ag(111) surface lattice constant.}
    \label{fig:CoPosErrors}
\end{figure}
}

\section{Code and data availability}
Code used to automate atom manipulation via the Createc software, analyze experimental data, theoretically model quantum corrals and create figures is provided at  \url{https://github.com/abekipnis/Kondo-Aalto} and makes use of the py\_createc Python package \cite{doi:10.1021/acsnano.1c06840}.

\bibliography{refs}